\documentclass[unsortedaddress,pre,reprint,showpacs]{revtex4-1}

\usepackage{amsmath,empheq}
\usepackage{bm}
\usepackage{amsfonts}
\usepackage{amsthm}
\usepackage{amssymb}
\usepackage{graphicx}
\usepackage{hyperref}
\usepackage{caption}
\usepackage{subcaption}
\usepackage{color}
\usepackage{soul}
	\newcommand{\ncd}{\newcommand}
	\ncd{\mrm}    {\mathrm}
	\ncd{\beq} {\begin{equation}}
	\ncd{\eeq} {\end{equation}}

	\def\p{{\bm p}}
	\def\q{{\bm q}}

        \def\z{{\bm{z}}}
        \def\l{{\bm{\lambda}}}

        \def\avg#1{\left< #1 \right>}

\begin{document}
	\title{Thermodynamic Cost for Classical Counterdiabatic Driving}
	\date{\today}

	\author{Alessandro Bravetti}
        \email{alessandro.bravetti@iimas.unam.mx}
        \affiliation{Instituto de Investigaciones en Matem\'aticas Aplicadas y en Sistemas, Universidad Nacional Aut\'onoma de M\'exico,
          Ciudad Universitaria, Ciudad de M\'exico 04510, Mexico}

	\author{Diego Tapias}
	\email{diego.tapias@nucleares.unam.mx }
        \affiliation{Departamento de F\'isica, Facultad de Ciencias, Universidad Nacional Aut\'onoma de M\'exico, Ciudad Universitaria, Ciudad de M\'exico 04510, Mexico}

	\begin{abstract}
Motivated by the recent growing interest about the thermodynamic cost of Shortcuts to Adiabaticity (STA), we consider the cost
of driving a classical system by the so-called Counterdiabatic Driving (CD). To do so, we proceed in three steps: first we review a general definition recently put forward in the literature
for the thermodynamic cost of driving a Hamiltonian system; then we provide a new  {complementary} definition of cost,  {which is of particular relevance} 
for cases where the average excess work vanishes; finally,
we apply our general framework to the case of CD. Interestingly, we find that  {in such case} our results are the exact classical counterparts of those reported in~\cite{funo2017universal}.
In particular we show that  {a universal trade-off between speed and cost for CD also exists in the classical case.}
 {To illustrate our points we consider the example of a time-dependent harmonic oscillator subject to different strategies of adiabatic control.}
	\end{abstract}


\maketitle

\section{Introduction}

{Shortcuts to Adiabaticity} (STA) is the design of nonadiabatic processes that reproduce in a finite time the same final state that would result from an adiabatic, infinitely slow, protocol.
Due to their importance in developing techniques for nanoengineering, such strategies have called the attention of the scientific community during the last years both
at the theoretical and at the experimental 
level~\cite{demirplak2003adiabatic,berry2009transitionless,ibanez2012multiple,jarzynski2013generating,del2013shortcuts,torrontegui2013shortcuts,couvert2008optimal,chen2010fast,schaff2011shortcut,bason2012high,
an2016shortcuts,martinez2016engineered}. 
Although the main practical motivation for STA is related to their quantum applications, 
there is a growing interest in understanding their classical counterparts, 
both because the classical scheme can be useful for new quantum strategies and because STA can be exploited for the design of protocols
that can speed up the convergence in Jarzynski's equality~\cite{jarzynski2013generating, deng2013boosting,deffner2014classical, jarzynski2017fast, okuyama2017quantum,acconcia2015degenerate,acconcia2015shortcuts}. 

One of the most noteworthy techniques for STA is Counterdiabatic Driving (CD).
In CD one uses an auxiliary Hamiltonian appropriately tailored so that the dynamics generated by the original Hamiltonian plus the auxiliary term 
preserves exactly the adiabatic invariant of the system~\cite{del2013shortcuts,jarzynski2013generating, deng2013boosting,deffner2014classical, jarzynski2017fast, okuyama2017quantum,sels2017minimizing}.
Moreover, since the auxiliary term vanishes at the beginning and at the end of the CD, one can show that the work distribution at the end of the protocol is the same as 
that of the bare adiabatic process~\cite{deng2013boosting}. 
Therefore it would seem that such driving effectively boosts the adiabatic dynamics at a finite rate without any extra cost
and that 
the duration of the CD process
can 
be pushed to zero with no a priori bound~\cite{del2014more,beau2016scaling},  {besides those set by quantum mechanics~\cite{bhattacharyya1983quantum}.}
This seems to be suspicious from a thermodynamic perspective. 
{Indeed,} recently there has been a surge of interest in trying to quantify the real cost of driving systems by CD 
and more in general by STA~\cite{zheng2016cost,deffner2017quantum, funo2017universal,campbell2017trade,abah2016performance,torrontegui2017energy}.
In particular, in~\cite{funo2017universal} a sensible definition of the thermodynamic cost for CD in the quantum case has been given, based on 
 {a universal trade-off between the speed and}
the excess of work fluctuations
at any intermediate time during the protocol, and it has been
shown that such cost 
 {can be further}
related to the geometric structure of the Hilbert space. 

In this work we revisit such results for classical systems and we set them into the more general perspective of the thermodynamic cost for driving a Hamiltonian system
out of equilibrium put forward in~\cite{sivak2012thermodynamic}.
Remarkably, we obtain that the most natural extension of the thermodynamic cost introduced in~\cite{sivak2012thermodynamic} when applied to CD provides the exact classical 
counterpart of the analysis in~\cite{funo2017universal},
 {including the existence of a universal bound on the speed of CD}. 
Moreover, as a by-product we obtain the classical equivalent of 
the geometric tensor proposed by Provost and Vallee~\cite{funo2017universal, provost1980riemannian}.
 {We conclude with two examples, the first one illustrating the bound for CD and the second one focusing on the importance of considering the excess of work fluctuations
also for STA different from CD.
} 

\section{Thermodynamic cost of driving}
In this section we briefly review the definition of the thermodynamic cost for driving a Hamiltonian system out of equilibrium given in~\cite{sivak2012thermodynamic} 
(see also~\cite{zulkowski2012geometry,zulkowski2013optimal,sivak2016thermodynamic})
in terms of the average excess work
and then we extend it to consider fluctuations of the excess work, in order to be able to deal with
cases where the average excess work vanishes, as it happens e.g.~in CD.

Let us consider a driven Hamiltonian system with Hamiltonian $H_0(\z,\l(t))$, where $\z=(\p, \q)$ denotes a point in phase space  and $\l(t)$ represents a set of time-dependent parameters
that can be externally moved according to a predetermined protocol.
One can define the conjugate forces and their deviations from the equilibrium values as
	\beq\label{DeltaX}
	{\bf X}:=\frac{\partial H_0}{\partial \l}  \quad {\rm and} \quad \Delta {\bf X}:={\bf X}-\avg{{\bf X}}_{\lambda}	\,.
	\eeq
Here the notation is the same as in~\cite{sivak2012thermodynamic}  (up to a sign in the first definition)
and $\avg{\ldots}_{\l}$ 
stands for the equilibrium average at fixed values of the control parameters $\l(t)$. 
Accordingly, the  (instantaneous) average power and the average power excess are
	\beq\label{P1}
	\avg{\mathcal{P}}_{\l}:=\dot{\l}\cdot\avg{{\bf X}}_{\l}  \quad {\rm and} \quad \avg{\mathcal{P}_{\rm ex}}_{\bf\Lambda}:=\dot{\l}\cdot \avg{\Delta {{\bf X}}}_{\bf\Lambda}\,,
	\eeq
where the subscript $\bf\Lambda$ denotes the average over the ensemble following the dynamics.
Using the standard definition of work~\cite{jarzynski1997nonequilibrium}, the average excess work can be calculated to be~\cite{sivak2012thermodynamic}
	\begin{eqnarray}
	\avg{W_{\rm ex}(t)}_{\bf\Lambda}&=&\avg{\int_{0}^{t}\dot\l\cdot \left(\frac{\partial H_{0}}{\partial \l}-\avg{\frac{\partial H_{0}}{\partial \l}}_{\l}\right)dt'}_{\bf\Lambda}\nonumber\\
	&=&\int_{0}^{t}\avg{\mathcal{P}_{\rm ex}}_{\bf\Lambda}dt'\,,\label{AvgWex}
	\end{eqnarray}
and thus 
	\beq\label{TDcost1}
	\avg{W_{\rm ex}}_{\tau}^{(1)}:=\tau^{-1}\int_{0}^{\tau}\avg{W_{\rm ex}(t)}_{\bf\Lambda}dt\,.
	\eeq
represents a good general definition for the thermodynamic cost of performing the nonequilibrium protocol on the system.

However, differently from~\cite{sivak2012thermodynamic}, here we want to consider not only the possibility of arbitrarily moving the parameters $\l$, but also
that of using auxiliary fields to control the evolution of the system over a predetermined path, as it happens in the case of STA.
As we will see, in such case the average excess work vanishes. Therefore it is important to generalize the above construction and define a thermodynamic
cost which applies to the case of vanishing
average excess work.
This is the task of the remainder of this section.
The most natural generalization of the cost~\eqref{AvgWex} is provided by considering the fluctuations of the excess work. 
From the definitions above it follows that
	\beq\label{P2}
	\avg{\mathcal{P_{\rm ex}}^{2}}_{\bf\Lambda}=\dot\lambda^{i}\dot\lambda^{j}\avg{\Delta X_{i}\Delta X_{j}}_{\bf\Lambda}
	\eeq
and accordingly one can compute
	\begin{eqnarray}
	\avg{W_{\rm ex}^{2}(t)}_{\bf\Lambda}&=&\int_{0}^{t}\int_{0}^{t}\avg{\mathcal{P}_{\rm ex}^{2}}_{\bf\Lambda}dt'dt''\nonumber\\
				&=&\int_{0}^{t}\int_{0}^{t}\dot\lambda^{i}\dot\lambda^{j}\avg{\Delta X_{i}\Delta X_{j}}_{\bf\Lambda}dt'dt''\,,\label{AvgWex2}
	\end{eqnarray}
together with the associated thermodynamic cost
	\beq\label{TDcost0}
	\avg{W_{\rm ex}}_{\tau}^{(2)}:=\tau^{-1}\int_{0}^{\tau}\sqrt{\avg{W_{\rm ex}^{2}(t)}_{\bf\Lambda}}dt\,.
	\eeq
This definition is completely general and applies to any system for which the combined action of the driving protocol and of the external fields
makes the average excess work vanish at any moment of time. In the following we consider one such case, i.e.~CD.

\section{The Cost of CD}
Let us now apply the above general discussion to the case of STA and in particular to CD.
The strategy of CD is to add an auxiliary term $H_1(\z, \l(t))$ to the bare Hamiltonian of the system, 
in such a way that the dynamics under the total Hamiltonian function $H_{\rm{CD}} = H_0 + H_1$ preserves the adiabatic invariant
	\beq
	\Omega(E,\bm\lambda) := \int d \bm z\, \Theta[E(\bm\lambda) - H_{0}(\bm z,\bm\lambda)]\,
	\label{shell}
	\eeq
exactly for all $t \in [0, \tau]$, with $\tau$ being the duration of the protocol~\cite{jarzynski2013generating, okuyama2017quantum}. 
From now on we consider only systems with one degree of freedom, as it is usual in the analysis of classical STA. 
In such case, it has been shown that $H_1(z, \l)$ can be written as 
	\beq
	H_1(z,\l) =  \dot\l \cdot {\bm \xi} = \dot{\lambda}^{i} \xi_{i}(z,\bm\lambda)\,,
	\label{h1}
	\eeq
 where $\bm{\xi}$ is the generator that 
 converts displacements in the space of parameters $\l \rightarrow \l + \delta \l$ 
 into displacements in the phase space $z \rightarrow z + \delta z$ 
 according to~\cite{jarzynski2013generating,deffner2014classical, okuyama2017quantum}
 	\beq
	\delta z=\delta \l \cdot\{z,{\bm\xi}\}\,.
	\eeq

The invariance of~\eqref{shell} may be restated in terms of the~\emph{adiabatic energy shell}
$\mathcal{E}(t) := \{z |  H_0(z, \l(t)) = E(t)\}$ as follows:
the CD guarantees that points starting on the same energy shell remain on the corresponding adiabatic energy shell during the evolution~\cite{jarzynski2017fast}. 
Considering this fact, it is natural to introduce a microcanonical measure associated with the dynamics under the CD control~\cite{jarzynski2013generating}
	\beq
	\langle \ldots \rangle_{E,\bm\lambda} := \frac{1}{\partial_{E} \Omega} \int \, d z \, \delta(E - H_{0}) \ldots\,,
	\label{micensemble}
	\eeq
where $E$ is the instantaneous energy of the adiabatic energy shell $\mathcal{E}(t)$. 
It follows that~\cite{jarzynski2013generating}
	\beq \label{xiavg}
	\avg{\bm \xi}_{ E,\bm\lambda} = 0\,,
	\eeq 
and that, for any initial distribution  of points over the phase space $\rho(z_0)$ that evolve under $H_{\rm{CD}}$, the mean value of an observable $f(z)$ at time $t$ is given by
	\beq
	\avg{f(z)}_{\bf\Lambda} := \int d z_0 \rho(z_0) \avg{f(z)}_{E,\bm\lambda} \,.
	\label{averages}
	\eeq
Using~\eqref{averages}, a natural definition of the instantaneous average of excess work is
	\beq
	\avg{W_{\rm{ex}}(t)}_{\bf\Lambda} := \avg{W(t) - W_{\rm{ad}}(t)}_{\bf\Lambda}\,,
	\label{workex}
	\eeq
where
	\beq
	{W(t)} = {H_{\rm{CD}}(z_{t},\bm\lambda_{t})-H_{0}(z_{0},\bm\lambda_{0})}
	\eeq
is the work done by the CD up to time $t$ and
	\beq
	{W_{\tiny\rm ad}(t)} = {H_{0}(z_{t},\bm\lambda_{t})-H_{0}(z_{0},\bm\lambda_{0})}\label{adW}
	\eeq
is the corresponding work done by the bare adiabatic protocol.
Note that in~\eqref{workex} we use the same notation as in~\eqref{AvgWex} because they coincide, as we now show.
From the definitions~\eqref{workex}--\eqref{adW}, one immediately arrives at
	\beq
	\avg{W_{\rm{ex}}(t)}_{\bf\Lambda} = \avg{H_1(z_t,\l_t)}_{\bf\Lambda} = \dot{\lambda}^i \avg{\xi_i}_{\bf\Lambda} \, .
	\label{meanex}
	\eeq
On the other side, starting from~\eqref{AvgWex} we have 
	\begin{eqnarray}
	\avg{W_{\rm{ex}}(t)}_{\bf\Lambda}
			&=&\int_{0}^{t}\dot\l\cdot \avg{\frac{\partial H_{\rm CD}}{\partial \l}-\avg{\frac{\partial H_{\rm CD}}{\partial \l}}_{\l}}_{\bf\Lambda}dt'\nonumber\\
			&=&\int_{0}^{t}\dot\l\cdot\avg{\frac{\partial H_{\rm CD}}{\partial \l}-\frac{\partial H_{0}}{\partial \l}}_{\bf\Lambda}dt'\nonumber\\
			&=&\avg{\int_{0}^{t}\dot\lambda^{a}\frac{\partial \dot\lambda^{i}{\xi_{i}}}{\partial \lambda^{a}}dt'}_{\bf\Lambda}\nonumber\\	
			&=&\dot{\lambda}^i\avg{\xi_i}_{\bf\Lambda}\,,\label{equivAvgW}
	\end{eqnarray}
where in the second equality we used~\eqref{xiavg} and  
{$\avg{\partial H_{0}/\partial\l}_{\l}=\avg{\partial H_{0}/\partial\l}_{\bf\Lambda}$},
which follows from observing that
 the dynamics with the CD forces each point
to follow the equilibrium trajectory of the bare system.

Clearly, from~\eqref{xiavg} and~\eqref{averages}, the above equations imply that on average the CD does no extra work with respect to the bare adiabatic protocol, as in the quantum case~\cite{funo2017universal}.

Since the average excess work vanishes, to study the cost of CD we now focus on the excess of work fluctuations according to~\eqref{AvgWex2}.
Repeating the steps in~\eqref{equivAvgW}, we arrive at the expression
	\begin{eqnarray}
	\avg{W_{\rm ex}^{2}(t)}_{\bf\Lambda}
				&=&\int_{0}^{t}\int_{0}^{t}\dot\lambda^{a}\dot\lambda^{b}\avg{\frac{\partial \dot\lambda^{i}{\xi_{i}}}{\partial \lambda^{a}}\frac{\partial \dot\lambda^{j}{\xi_{j}}}{\partial \lambda^{b}}}_{\bf\Lambda}dt'dt''\nonumber\\					&=&\avg{\int_{0}^{t}\int_{0}^{t}\dot\lambda^{a}\dot\lambda^{b}\frac{\partial \dot\lambda^{i}{\xi_{i}}}{\partial \lambda^{a}}\frac{\partial \dot\lambda^{j}{\xi_{j}}}{\partial \lambda^{b}}dt'dt''}_{\bf\Lambda}\nonumber\\	
				&=&\dot\lambda^{i}\dot\lambda^{j}\avg{\xi_{i}\xi_{j}}_{\bf\Lambda}\,.	\label{AvgWex2bis}
	\end{eqnarray}
This increase in the work fluctuations is the price to be paid in the case of CD and it has a thermodynamic cost quantified by~\eqref{TDcost0}.	
Indeed, in the following we show that this definition of the thermodynamic cost is exactly the classical counterpart of the one proposed in~\cite{funo2017universal}.

Following~\cite{funo2017universal}, we now define the difference in the instantaneous work fluctuations as
	\beq
	\delta(\Delta W)^2:={\rm Var}[W(t)] - {\rm Var}[W_{\tiny\rm ad}(t)] \, .
	\eeq
where `Var' stands for the variance. Expanding each term on the right hand side and 
then using~\eqref{meanex}, we arrive at the following result
	\begin{align}
	\delta(\Delta W)^2
	&=\avg{W(t)^2}_{\bf\Lambda} -  \avg{W(t)}_{\bf\Lambda}^2 - \avg{W_{\tiny\rm ad}(t)^{2}}_{\bf\Lambda} +  \avg{W_{\tiny\rm ad}(t)}_{\bf\Lambda}^2 \notag \\
	&= \avg{H_{\tiny\rm CD}(z_{t},\bm\lambda_{t})^2 - H_{0}(z_{t},\bm\lambda_{t})^2}_{\bf\Lambda} \notag \\
	&= \avg{H_{1}(z_{t},\bm\lambda_{t})^2 }_{\bf\Lambda} \nonumber\\
	&=\dot\lambda^{i}\dot\lambda^{j}\avg{\xi_{i}\xi_{j}}_{\bf\Lambda}\,,\label{Ksq}
	\end{align}
that is, the instantaneous modifications in the work fluctuations due to the CD coincide with the second
moment of the excess work (cf.~eq.~\eqref{AvgWex2bis}).
Moreover, from~\eqref{Ksq} and~\eqref{averages}, we obtain
	\beq\label{variance}
	\delta(\Delta W)^2
	=\int d z_{0}\rho(z_{0})\avg{\xi_{i}\xi_{j}}_{E,\bm\lambda}\dot{\lambda}^{i}\dot{\lambda}^{j}\,,
	\eeq
and from comparison with~\cite{funo2017universal} we conclude that
	\beq\label{newg}
	g_{ij}^{E}:=\avg{\xi_{i}\xi_{j}}_{E,\bm \lambda}
	\eeq
is the classical equivalent of the quantum geometric tensor
	\beq
	{g_{ij}^{(n)}=\langle n(t)|\hat{\xi}_{i}^{\dagger} \hat{\xi}_{j}| n(t)\rangle}\,.
	\eeq 
 {This metric induces a natural distance between the initial and final statistical states of the system  under the CD dynamics, $\rho(z_0)$ and $\rho(z_\tau)$, given by
	\beq\label{length}
	l(\rho(z_0),\rho(z_\tau)):=\int_{0}^{\tau}\sqrt{\delta(\Delta W)^2}\,d t\,.
	\eeq
Notice that such distance is well defined because $\delta(\Delta W)^2\geq 0$.
}

\section{A classical Speed Limit for CD}
 {
In this section we use the results in the previous section to show that there exists a universal trade-off between speed and cost for CD in the classical case.
}

 {
From~\eqref{TDcost0}, \eqref{AvgWex2bis}, \eqref{Ksq} and~\eqref{length} it follows that
	\beq\label{TDcostnew}
	\avg{W_{\rm ex}}_{\tau}^{(2)}=\frac{1}{\tau}\,l(\rho(z_0),\rho(z_\tau))=\avg{\delta\Delta W}_{\tau}\,,
	\eeq
where $\delta\Delta W:=\sqrt{\delta(\Delta W)^{2}}$.
Therefore we have expressed the thermodynamic cost~\eqref{TDcost0} in terms of the statistical length between the initial and final distribution in phase space.
Furthermore, considering that $\delta(\Delta W)^{2}=\avg{H_{\rm CD}^{2}}_{\bf\Lambda}-\avg{H_{0}^{2}}_{\bf\Lambda}$ (cf.~equation~\eqref{Ksq}), 
and noting that $\avg{H_{0}^{2}}_{\bf\Lambda}\geq\avg{H_{0}}_{\bf\Lambda}^{2}=\avg{H_{\rm CD}}_{\bf\Lambda}^{2}$,
we obtain the following inequality
	\beq
	\delta(\Delta W)^{2}\leq \avg{H_{\rm CD}^{2}}-\avg{H_{\rm CD}}^{2}= {\rm Var}[H_{\rm CD}]
	\eeq
and hence, using~\eqref{length}, we get
	\beq\label{bound}
	\tau\geq\frac{l(\rho(z_0),\rho(z_\tau))}{\avg{\rm{Std} [H_{\rm CD}]}_{\tau}} \,,
	\eeq
where `Std' is the standard deviation.
This result provides the same bound on the duration of the process as the one obtained in~\cite{funo2017universal} 
(although in~\cite{funo2017universal} it is further reduced with the use of the Bures distance in the Hilbert space). 
The remarkable point about our derivation of such relation is the fact that we showed explicitly that it does not
depend on any quantum property. We conclude that this bound depends solely on the nature of the counterdiabatic control and it sets a  limit on the duration of a CD
exactly in the same way as its quantum counterpart.
}

\section{Examples}
 {
In this section we illustrate the above results using a paradigmatic example, namely the driven harmonic oscillator,
with bare Hamiltonian
	\beq\label{harmosc}
	H_0(p,q, \omega(t)) = \dfrac{p^2}{2} + \omega^2(t)\dfrac{q^{2}}{2} \, .
	\eeq
We analyze first the {CD}, showing that the bound~\eqref{bound} forbids drivings at any speed, and then we consider a different type of driving, which is not CD
{but still}
 realizes a STA for this system. We use this latter case to 
 argue for the need of taking into account the cost provided by the average excess of work fluctuations also in cases different from CD. 
}

\subsection{With CD}\label{subsec:CD}
 {
The auxiliary Hamiltonian~\eqref{h1} in this case reads~\cite{deffner2014classical}
	\beq
	H_1(p,q, \omega(t)) = -\frac{p q}{2 \omega(t)} \dot{\omega}(t) \, .
	\eeq
For a general protocol $\omega(t)$ we have that the adiabatic energy shell is described by the ellipse
	\beq\label{Eshell}
	 \mathcal{E}(t) = \left\{(p,q): \dfrac{p^2}{2} + \dfrac{\omega(t)^2}{2} q^2 =  \frac{E_0}{\omega_0} \omega(t)\right\} \, ,
	\eeq
where in the last equality we used the adiabatic invariant for the harmonic oscillator~\cite{goldstein2014classical}.  
In this case there is only one parameter $\omega$ and thus there is only one metric component in~\eqref{newg},
which can be calculated at any time $t$ using the energy shell~\eqref{Eshell}, to obtain
	\beq
	g_{11} = 
	\frac{E_0^2 \left(2 \left(\omega^4-\omega^2+1\right) F\left(X\right)-\left(\omega^2+1\right)
	   K\left(X\right)\right)}{15 \omega^2 \left(\omega^2-1\right)^2 F\left(X\right)}\,.
	\eeq
Here $\omega:=\omega(t)$, $X:=1-{1}/{\omega^2}$, $F\left(X\right)$ and $K\left(X\right)$ are the complete elliptic integral of the second and first kind respectively,
 and we are using $\omega_{0}=1$ throughout the rest of the paper.
Considering an initial canonical distribution with $\beta = 1$, together with the protocol
	\beq\label{protocol1}
	\omega(t) = 1 + 20 s^3  - 30 s^4 + 12s^5\,,
	\eeq 
(cf.~\cite{funo2017universal}),
where $s:=t/\tau$, we can compute numerically the instantaneous excess of work fluctuations~\eqref{variance} 
and the thermodynamic cost~\eqref{TDcostnew} for different values of $\tau$. The results are displayed in Fig.~\ref{cost}.
\begin{figure}[h!]
  \includegraphics[width=.23\textwidth]{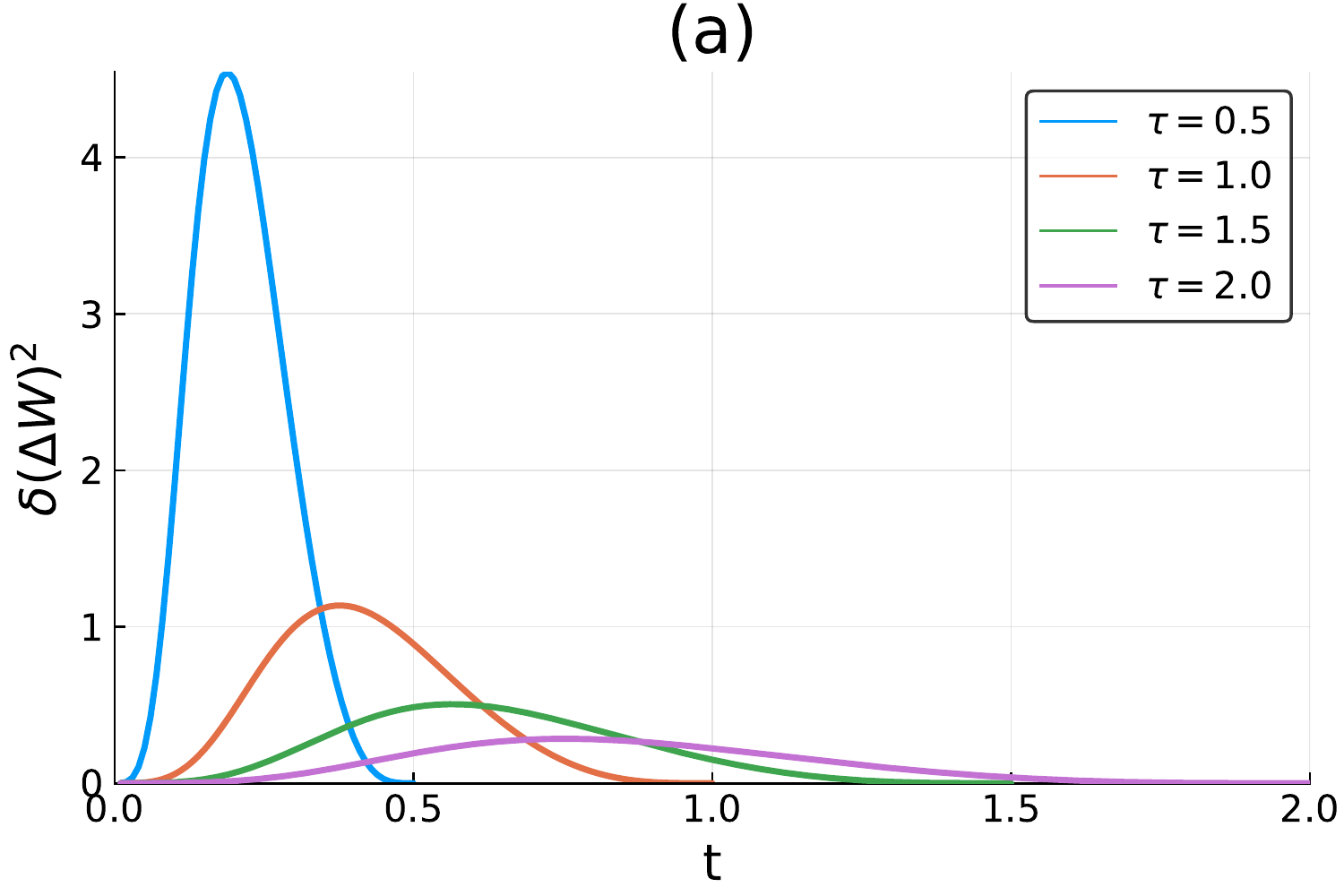}
 \includegraphics[width=.23\textwidth]{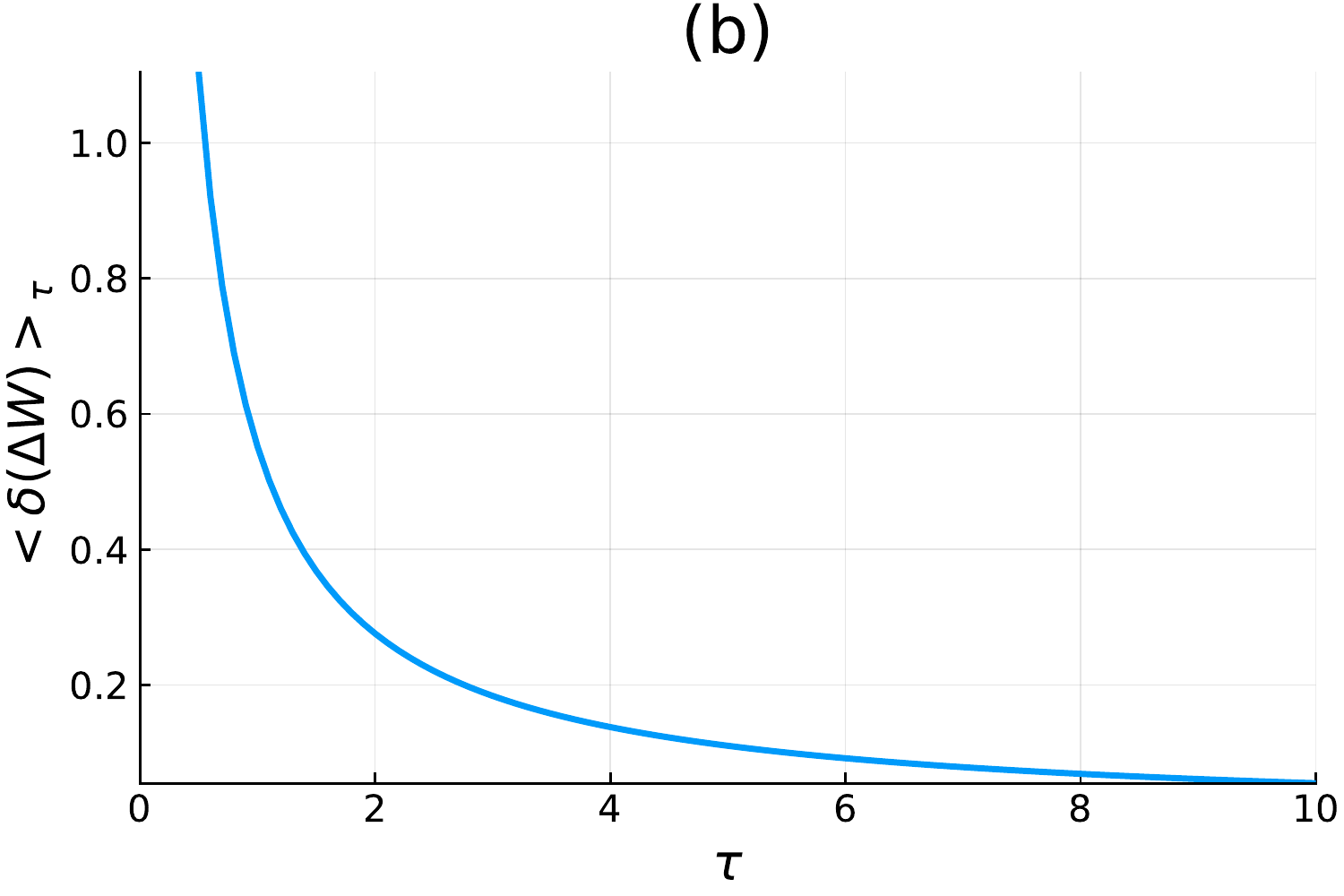}
\caption{
Instantaneous excess of work fluctuations (a) and thermodynamic cost (b) for different durations $\tau$ of the protocol~\eqref{protocol1} for a time-dependent harmonic oscillator.}
\label{cost}
\end{figure}
As expected, the shorter the protocol the larger the instantaneous excess of work fluctuations; 
this affects the thermodynamic cost of the CD, which diverges in the limit $\tau \rightarrow 0$, in agreement with~\eqref{bound}.
}

\subsection{With STA from linear response theory}
 {
In~\cite{acconcia2015degenerate}, the authors found a whole class of degenerate protocols which realize a STA for the harmonic oscillator~\eqref{harmosc}
under weak driving, obtained from linear response theory (see also~\cite{acconcia2015shortcuts}).
The relevance of such protocols here is the fact that, unlike CD, they have $\avg{W_{\rm ex}(t)}_{\bf\Lambda}\neq 0$ for $0<t<\tau$, while they
all have $\avg{W_{\rm ex}(\tau)}_{\bf\Lambda}=0$ for appropriate values of $\tau$. Thus we can use these examples to illustrate the importance
of comparing the excess of work fluctuations during the protocol with the mean excess work, by showing that there are further cases (beside CD) 
in which the former term cannot be neglected.
We remark that this is something one would intuitively expect for small systems.
We consider here in particular two protocols: the linear one
	\beq\label{linprotocol}
	\omega^2_{\rm lin}(t) = 1 + \frac{1}{10}s\,,
	\eeq
and the sine protocol
	\beq\label{sineprotocol}
	\omega^2_{\rm sin}(t) = 1 +  \frac{1}{10}\left(s+{\rm sin}\left(2\pi s\right)\right)\,,
	\eeq
(cf.~\cite{acconcia2015degenerate}). Taking a canonical initial distribution with $\beta=1$ as in the CD case (cf.~Sec.~\ref{subsec:CD})
and using values of $\tau$ for which such protocols effectively provide a STA ($\tau=\pi$ for the linear case and $\tau\approx 1.2$ for the sine protocol), we can compute
numerically both the thermodynamic cost stemming from the average excess work at any intermediate time -- equation~\eqref{TDcost1} -- and its analogue stemming from
the excess of work fluctuations, equation~\eqref{TDcost0}.
In Table~\ref{costtable} we show the results. In both cases $\avg{W_{\rm ex}}_{\tau}^{(2)}$ is greater than $\avg{W_{\rm ex}}_{\tau}^{(1)}$
by two orders of magnitude. Therefore we consider that estimates of the thermodynamic cost of a protocol based on the average work excess alone cannot be fully precise in such cases. 
\begin{table}[h!]
\begin{tabular}{|c|c|c|}
\hline 
Protocol & 	$\avg{W_{\rm ex}}_{\tau}^{(1)}$ & $\avg{W_{\rm ex}}_{\tau}^{(2)}$  \\ 
\hline 
$\omega^2_{\rm lin}$ & $1.27 \times 10^{-4}$ & $2.75 \times 10^{-2}$ \\ 
\hline 
$\omega^2_{\rm sin}$ & $4.78 \times 10^{-4}$  & $5.84 \times 10^{-2}$ \\ 
\hline 
\end{tabular}
\caption{
Comparison between the thermodynamic costs based on the first two moments of average excess work
for the protocols~\eqref{linprotocol} and \eqref{sineprotocol}.}
\label{costtable}
\end{table}
This aspect has been somehow overlooked in the literature and further research in this direction is required
in all cases where similar controls are proposed.
}

\section{Conclusions}
In this work we have formulated a general framework to analyze the thermodynamic cost
of driving a Hamiltonian system by using the combined action of moving its parameters and possibly also adding
{extra} fields. Interestingly, in the particular case of CD our results provide the classical counterpart of the findings in~\cite{funo2017universal}. 
Besides, the metric~\eqref{newg} {is} completely new in the classical case  {(see also the related discussion in~\cite{kolodrubetz2013classifying})}. 
For instance,
we remark that this tensor is different from the one proposed in~\cite{sivak2012thermodynamic},
because
there the metric quantifies the average excess power,
which vanishes in the case of CD. 
 {In addition, we showed that there exists a speed limit~\eqref{bound} for CD stemming from~\eqref{newg}, which provides a bound on the speed
at which CD can be performed, exactly in the same way as in the quantum case. However here the analysis is completely classical. Based on this argument,
we argued} that~\eqref{newg} is  {a} relevant object for the analysis of the thermodynamic cost of CD, and that more in general one should consider 
 {both the cost based on}
the average excess work and 
 {that based on the average excess of work fluctuations according to the definitions~\eqref{TDcost1} and~\eqref{TDcost0} above.
Although we worked here only with systems with one degree of freedom and the extension of all the results to interacting many-body systems 
is not direct (see e.g.~the discussion in~\cite{weinberg2017adiabatic}),
we expect that these tools will provide useful complementary information for assessing the thermodynamic cost
of general driving protocols {for} classical systems.}

\section{Acknowledgements}
The authors would like to thank A del Campo, K Funo, and M Ueda for their valuable comments. 
AB is funded by a DGAPA--UNAM postdoctoral fellowship. DT acknowledges financial support from
Conacyt (CVU~442828).

\bibliography{20170912v3}

\end{document}